\documentclass{elsart1p}

\usepackage{graphicx}

\usepackage{amssymb,longtable,amsmath}

\begin{document}

\begin{frontmatter}



\title{Charged-Particle Thermonuclear Reaction Rates:\\ I. Monte Carlo Method and Statistical Distributions}


\author{R. Longland}, \author{C. Iliadis}, \author{A. E. Champagne}, \author{J.~R. Newton}, \author{C. Ugalde}
\address{Department of Physics and Astronomy, University of North Carolina, Chapel Hill, NC 27599-3255, USA; Triangle Universities Nuclear Laboratory, Durham, NC 27708-0308, USA}
\author{A. Coc}
\address{Centre de Spectrom\'etrie Nucl\'eaire et de Spectrom\'etrie de Masse (CSNSM), UMR 8609, CNRS/IN2P3 and Universit\'e
        Paris Sud 11, B\^atiment 104, 91405 Orsay Campus, France}
\author{R. Fitzgerald}
\address{National Institute of Standards and Technology, 100 Bureau Drive, Stop 8462, Gaithersburg, MD 20899-8462, USA}

\begin{abstract}
A method based on Monte Carlo techniques is presented for evaluating thermonuclear reaction rates. We begin by reviewing commonly applied procedures and point out that reaction rates that have been reported up to now in the literature have no rigorous statistical meaning. Subsequently, we associate each nuclear physics quantity entering in the calculation of reaction rates with a specific probability density function, including Gaussian, lognormal and chi-squared distributions. Based on these probability density functions the total reaction rate is randomly sampled many times until the required statistical precision is achieved. This procedure results in a median (Monte Carlo) rate which agrees under certain conditions with the commonly reported recommended ``classical" rate. In addition, we present at each temperature a low rate and a high rate, corresponding to the 0.16 and 0.84 quantiles of the cumulative reaction rate distribution. These quantities are in general different from the statistically meaningless ``minimum" (or ``lower limit") and ``maximum" (or ``upper limit") reaction rates which are commonly reported. Furthermore, we approximate the output reaction rate probability density function by a lognormal distribution and present, at each temperature, the lognormal parameters $\mu$ and $\sigma$. The values of these quantities will be crucial for future Monte Carlo nucleosynthesis studies. Our new reaction rates, appropriate for {\it bare nuclei in the laboratory}, are tabulated in the second paper of this series (Paper II). The nuclear physics input used to derive our reaction rates is presented in the third paper of this series (Paper III). In the fourth paper of this series (Paper IV) we compare our new reaction rates to previous results.
\end{abstract}


\end{frontmatter}

\section{Introduction}\label{intro}
The most influential charged-particle thermonuclear reaction rate evaluations of the 20th century were published by Fowler and collaborators in a series of several papers, with the latest being published in 1988 \cite{CF88}. The latter work provided compiled rates in tabular and in analytical format for 128 proton- and $\alpha$-particle induced reactions on A=1 to 30 nuclei. About a decade later, a new reaction rate evaluation by the NACRE collaboration \cite{Ang99} updated many of the previously published results. The NACRE evaluation contains the rates of 86 reactions on A=1 to 28 nuclei in tabular and analytical format. It represented a major improvement, not only by including newly available nuclear physics input, but it provided for the first time: (i) estimates of reaction rate uncertainties at each temperature in tabular format, and (ii) most of the nuclear data and the associated references used to derive the reaction rates. Another evaluation was published in 2001 by Iliadis and collaborators \cite{Ili01}. These authors  provided 55 reaction rates involving A=20 to 40 target nuclei in tabular format. They presented reaction rate uncertainties in graphical format and most of the nuclear physics input used to compute the rates. The two major innovations of the latter work were: (i) an extension of the rate evaluation effort to reactions involving radioactive target nuclei, and (ii) the normalization of many resonance strengths to a ``backbone" of selected and carefully measured standard strengths.

The fast progress seen in the field of nuclear astrophysics over the past few years warrants a new reaction rate evaluation. The original aim was to publish in a short paper the reaction rates that were recently updated by one of us (CI) while working on a textbook \cite{Ili07} and thus to make them available to the community of stellar modelers. However, it became quickly obvious that there are significant problems in all previously published reaction rate evaluations when the results are confronted with some basic ideas of statistics: what is the statistical meaning of published reaction rates and their uncertainties? Do the published rate uncertainties represent standard deviations of Gaussian distributions or do they perhaps correspond to some other coverage probability? What is the precise meaning of published ``upper" and ``lower" limits? And, finally, how can published reaction rate uncertainties be used in the calculations they are mainly intended for, that is, in stellar models? 

We argue here that reaction rates from previously published evaluations have no precise statistical meaning. The present work is part of a series of four papers on a new evaluation of charged-particle thermonuclear reaction rates on A=14 to 40 target nuclei. In the first paper, referred to as Paper I, we present a method based on Monte Carlo techniques of estimating statistically meaningful reaction rates and their associated uncertainties\footnote{It is regrettable that the terms {\it uncertainty} and {\it error} are used interchangeably in the nuclear astrophysics literature. According to the {\it ISO Guide to the Expression of Uncertainty and Measurement (GUM)} \cite{GUM,NIST} these expressions ``...are not synonyms, but represent completely different concepts; they should not be confused with one another or misused...". The  {\it uncertainty} is defined as a ``parameter, associated with the result of a measurement, that characterizes the dispersion of the values that could reasonably be attributed to the measurand". Uncertainty of measurement comprises, in general, many components; some of these may be evaluated from the statistical distribution of the results of series of measurements and can be characterized by experimental standard deviations; other components, which also can be characterized by standard deviations, are evaluated from assumed probability distributions based on experience or other information. On the other hand, if we use the term {\it error} in connection with a reaction rate, it means that we think the rate is wrong since perhaps a correction for some systematic effect was disregarded.}. Paper II contains our numerical results in tabular format, while in Paper III we provide the complete nuclear physics data input used to derive our new reaction rates. In Paper IV we compare our new reaction rates to previous results.
 
The aim of the present work is to evaluate and compile charged-particle thermonuclear reaction rates for A=14 to 40 nuclei on a grid of temperatures ranging from T=0.01 GK to 10 GK. These reaction rates are assumed to involve {\it bare nuclei in the laboratory}. For use in stellar model calculations, the results presented here must be corrected, if appropriate, for (i) electron screening at elevated densities, and (ii) thermal excitations of the target nucleus at elevated temperatures. Although we occasionally used results from nuclear theory, the present reaction rates are overwhelmingly based on {\it experimental nuclear physics information}. Only in exceptional situations, for example, when a nuclear property had not been measured yet, did we resort to nuclear theory.

Paper I is organized as follows. In Sec. \ref{formalism} we present the formalism and the expressions used for computing reaction rates. The commonly employed and accepted procedure of estimating reaction rates and their associated uncertainties is briefly presented in Sec. \ref{classicalrates}. We refer to all results derived from this method, including those presented in Refs. \cite{CF88,Ang99,Ili01}, as ``classical reaction rates". It will become obvious that there are major problems from the statistics point of view with this method. Statistical distributions are briefly reviewed in Sec. \ref{statdistr} in order to provide a basis for the following discussion. Our method of estimating reaction rates, which is based on Monte Carlo techniques, is presented in Sec. \ref{MonteCarlo}. We will refer to the new results as ``Monte Carlo reaction rates". A summary and suggestions for future work are given in Sec. \ref{summary}.
%
\section{Reaction rate formalism}\label{formalism}
A recent discussion of the formalism can be found in Iliadis \cite{Ili07}. Here we summarize the most important results. In this section, all energies refer to the center-of-mass coordinate system. The total laboratory thermonuclear rate (in units of cm$^3$~mol$^{-1}$~s$^{-1}$) for a reaction involving two nuclei 0 and 1 in the entrance channel at a given temperature $T$ is given by 
\begin{equation}
N_A\langle \sigma v \rangle_{01} = \frac{3.7318\cdot 10^{10}}{T_9^{3/2}} \sqrt{\frac{M_0 + M_1}{M_0 M_1}} \int_0^\infty E\,\sigma(E)\,e^{-11.605\,E/T_9}\,dE \label{generalrate}
\end{equation}
where the center-of-mass energy $E$ is in units of MeV, the temperature $T_9$ is in GK ($T_9 \equiv T/10^9$~K), the atomic masses $M_i$ are in u and the cross section $\sigma$ is in b ($1~\mathrm{b} \equiv 10^{-24}~\mathrm{cm}^2$); $N_A$ denotes the Avogadro constant. Thus the reaction rate is determined by the absolute magnitude and the energy dependence of the nuclear reaction cross section $\sigma(E)$. Based on the energy-dependence of $\sigma(E)$, a number of different specialized expressions and procedures can be derived for certain contributions to the total reaction rate. These contributions will be discussed in the following.
%
\subsection{Nonresonant reaction rates}\label{nonresrates}

Nonresonant cross sections vary smoothly with energy and are usually converted into the astrophysical S-factor, defined by
\begin{equation}
S(E) \equiv E\,e^{2\pi\eta}\,\sigma(E)	\label{sfactor}
\end{equation}
This definition removes the 1/E dependence of nuclear cross sections and the s-wave Coulomb barrier transmission probability $e^{-2\pi\eta}$ (that is, the Gamow factor) from the cross section and yields a quantity, $S(E)$, that depends only weakly on energy. The Sommerfeld parameter $\eta$ is numerically given by
\begin{equation}
2\pi\eta  = 0.989510\,Z_0 Z_1 \sqrt{\frac{M_0 M_1}{M_0 + M_1}\frac{1}{E}}     \label{Sommerfeld}
\end{equation}
where $Z_i$ are the charges of nuclei 0 and 1. For a weak energy dependence of the S-factor, substitution of Eq. (\ref{sfactor}) into Eq. (\ref{generalrate}) yields an integrand whose energy dependence is dominated on the low-energy side by the penetrability through the Coulomb barrier and on the high-energy side by the Maxwell-Boltzmann distribution of the interacting nuclei. The integrand, which is referred to as Gamow peak, represents the energy range of effective stellar burning at a given temperature. The location of the maximum, $E_0$, and the (Gaussian approximation) 1/e width, $\Delta E_0$, of the Gamow peak (in units of MeV) are given by
\begin{equation}
E_0  = 0.1220\left(Z_0^2 Z_1^2 \frac{M_0 M_1}{M_0 + M_1}\,T_9^2 \right)^{1/3},~~~
\Delta E_0 = 0.2368\left(Z_0^2 Z_1^2 \frac{M_0 M_1}{M_0 + M_1}\,T_9^5\right)^{1/6}  	\label{gamow}
\end{equation}
The area enclosed between the 1/e points of a Gaussian amounts to 84\%. If the S-factor can be approximated by a polynomial,
\begin{equation}
S(E) \approx  S(0) + S'(0)E + \frac{1}{2} S''(0)E^2      \label{Sfactorpoly}
\end{equation}
where the primes indicate derivatives with respect to $E$, then the nonresonant reaction rate can be obtained from the relations
\begin{equation}
N_A \langle \sigma v \rangle_{\mathrm{nr}}  =  \frac{4.339 \cdot 10^{8}}{Z_0 Z_1}\, \frac{M_0 + M_1}{M_0M_1} \,S_{\mathrm{eff}} \, e^{-\tau} \tau^2       \label{nonresrateexpr}
\end{equation}
\begin{equation}
\tau = 4.2487\left(Z_0^2 Z_1^2 \frac{M_0 M_1}{M_0 + M_1}\frac{1}{T_9}\right)^{1/3}      \label{numericaltau}
\end{equation}
\begin{equation}
S_{\mathrm{eff}}  = S(0)\left[1 + \frac{5}{12\tau} + \frac{S'(0)}{S(0)}\left(E_0 + \frac{35}{36}kT\right) + \frac{1}{2}\frac{S''(0)}{S(0)}\left(E_0^2 + \frac{89}{36}E_0 kT\right)\right]     \label{Sfactorterm}
\end{equation}
with $S_{\mathrm{eff}}$ in Eq. (\ref{nonresrateexpr}) given in units of MeV$\cdot$b; $k$ denotes the Boltzmann constant. The nonresonant reaction rate expression is frequently multiplied by a cutoff factor 
\begin{equation}
f_{\mathrm{cutoff}} \approx e^{-(T_9/T_{9,\mathrm{cutoff}})^2} \label{cutoff}
\end{equation}
where $T_{9,\mathrm{cutoff}}$ corresponds to that temperature at which the Gamow peak starts to shift to energies at which the $S$-factor expansion of Eq. (\ref{Sfactorpoly}) becomes inaccurate.
%
\subsection{Narrow-resonance reaction rates}\label{narresrates}
The cross section of an isolated resonance can be described by the one-level Breit-Wigner formula (see later). A resonance can be considered as narrow if the partial widths and the Maxwell-Boltzmann factor $e^{-E/kT}$ are approximately constant over the total width of the resonance. The narrow-resonance reaction rate is then obtained by substitution of the Breit-Wigner formula into Eq. (\ref{generalrate}). The result is
\begin{equation}
N_A \langle {\sigma v} \rangle_{\mathrm{r}} = \frac{1.5399 \cdot 10^{11}}{T_9^{3/2}}\left( \frac{M_0 + M_1}{M_0 M_1 }\right)^{3/2}\sum\limits_i (\omega \gamma)_i e^{-11.605\,E_i/T_9}    \label{narresrateexpr}
\end{equation}
where the incoherent sum is over all narrow resonances $i$. In this expression, the resonance energies
$E_i$ and resonance strengths $(\omega \gamma)_i$ are both in units of MeV. In terms of resonance properties, the resonance strength is defined as 
\begin{equation}
\omega\gamma = \frac{2J+1}{(2j_0+1)(2j_1+1)} \frac{\Gamma_a \Gamma_b}{\Gamma}   \label{resstrength}
\end{equation}
with $J$, $j_0$, $j_1$ the spins of resonance, projectile and target nucleus, respectively, and $\Gamma_a$, $\Gamma_b$, $\Gamma$ the partial widths for the entrance and exit channel, and the total resonance width (that is, the sum of all partial widths, $\Gamma=\Gamma_a+\Gamma_b+...$) at the resonance energy, respectively. In a reaction with only two energetically allowed channels, one finds frequently for low-energy resonances that $\Gamma_a \ll \Gamma_b$. In this case, $\Gamma \approx \Gamma_b$ and thus $\omega\gamma \approx \omega \Gamma_a$. Note that Eq. (\ref{narresrateexpr}) contains the value of the Maxwell-Boltzmann distribution at $E_i$ and, consequently, this expression takes only the reaction rate contribution at the resonance energy into account.
%
\subsection{Broad-resonance reaction rates}\label{numintrates}
There are mainly two situations where Eq. (\ref{narresrateexpr}) becomes inaccurate for calculating the reaction rate contribution of a resonance. First, if a resonance is sufficiently broad the Maxwell-Boltzmann factor $e^{-E/kT}$ and the partial widths $\Gamma_i$ may vary with energy over the width of the resonance, leading to a breakdown of the main assumption used in deriving the narrow-resonance rate formula. Second, suppose that the temperature is gradually decreased such that the Gamow peak shifts away from a given resonance. At some point the rate contribution from the region around the resonance energy will then become negligible compared to the contribution arising from the Gamow peak (that is, from the wing of the resonance). The latter contribution is obviously not considered in Eq. (\ref{narresrateexpr}). As a rule of thumb, if the resonance energy falls outside the range $E_0 \pm 2\Delta E_0 $ then, even for a narrow resonance, the rate contribution from the resonance wing must be taken into account. 

In both of these situations the reaction cross section can be described by the one-level Breit-Wigner formula. For the cross section (in units of b) of a resonance located at energy $E_r$ we find
\begin{equation}
\sigma_{\mathrm{BW}}(E) = 0.6566 \frac{\omega}{E} \frac{M_0+M_1}{M_0 M_1} \frac{\Gamma_a(E)\Gamma_b(E + Q-E_f)}{(E_r - E)^2 + \Gamma(E)^2/4}	\label{breitwignercrosssection}
\end{equation}
where all energies and widths are in units of MeV; $\omega$ is the spin factor of Eq. (\ref{resstrength}), $Q$ is the reaction Q-value, and $E_f$ is the energy of the final state in the residual nucleus. In this expression and throughout this work the energy-dependent partial widths denote ``observed" rather than ``formal" quantities \cite{Lan58}. The particle partial width for a given level $\lambda$ and channel $c$ is defined by
\begin{equation}
\Gamma_{\lambda c} = 2 P_c \gamma_{\lambda c}^2 = 2 \frac{\hslash^2}{mR^2} P_c \theta_{\lambda c}^2    \label{partpartwidth}
\end{equation}
with $m=m_0m_1/(m_0+m_1)$ the reduced mass, $P_c$ the penetration factor, $\gamma_{\lambda c}^2$ the reduced width and $\theta_{\lambda c}^2$ the dimensionless reduced width; for the channel radius we chose the commonly used prescription $R=1.25(A_0^{1/3} + A_1^{1/3})$ fm, where the $A_i$ denote (integer) mass numbers of the interacting nuclei. The dimensionless reduced width for a single-particle channel can be parametrized as 
\begin{equation}
\theta_{\lambda c}^2= C^2S \theta_{p c}^2 \label{iliadisexpr}
\end{equation}
where $C$, $S$ and $\theta_{p c}^2$ denote an isospin Clebsch-Gordan coefficient, the single-particle spectroscopic factor and the dimensionless single-particle reduced width, respectively. Calculated values of $\theta_{p c}^2$ for protons can be found in Iliadis \cite{Ili97}. The partial width for a particular $\gamma$-ray transition is given by
\begin{equation}
\Gamma_{\gamma}(\overline{\omega} L) = \frac{8\pi (L+1)}{L[(2L+1)!!]^2} \left(\frac{E_\gamma}{\hslash c}\right)^{2L+1} B(\overline{\omega}L)      \label{gammapartwidth}
\end{equation}
with $E_\gamma$ and $L$ the energy and multipolarity of the radiation, respectively; $\overline{\omega}$ denotes either electric (E) or magnetic (M) radiation and the double factorial is defined as $(2L + 1)!! \equiv 1\cdot 3\cdot 5\cdot \ldots \cdot (2L + 1)$. The quantity $B(\overline{\omega} L)$ is the reduced transition probability. Note that in general the quantities $\Gamma_a$ and $\Gamma_b$ in Eq. (\ref{breitwignercrosssection}) represent, even for a transition to a specific final state, sums over different components of $\Gamma_{\lambda c}$ or $\Gamma_{\gamma}(\overline{\omega} L)$. For example, orbital angular momenta of $\ell$ and $\ell + 2$ may contribute to a particle partial width, or $\gamma$-ray multipolarities of M1 and E2 may contribute to a $\gamma$-ray partial width. When the partial widths at the resonance energy $E_r$ are known and if one value of $\ell$ or $\overline{\omega} L$ dominates the total particle or $\gamma$-ray partial width, it is usually more reliable to use instead of Eqs. (\ref{partpartwidth}) and (\ref{gammapartwidth}) the scaling relations
\begin{equation}
\Gamma_{\lambda c}(E) =  \Gamma_{\lambda c}(E_r) \frac{P_c(E)}{P_c(E_r)}   \label{scalingpart}
\end{equation}
\begin{equation}
\Gamma_{\gamma}(\overline{\omega} L, E) = \Gamma_{\gamma}(\overline{\omega} L, E_r) \left( \frac{E+Q-E_f}{E_r+Q-E_f} \right)^{2L+1}   \label{scalinggam}
\end{equation}
The energy $E$ in Eq. (\ref{scalingpart}) refers to the total kinetic energy in the particle channel: for the entrance particle channel, $E$ denotes the center-of-mass bombarding energy, while for an exit particle channel one has to replace $E$ by $E'=E+Q-E_f$. 

When a resonance cross section is given by Eq. (\ref{breitwignercrosssection}), no simple analytical reaction rate formula can be derived mainly because the Coulomb wave functions which determine the penetration factor $P_c$ must be evaluated numerically. The reaction rates for ``broad" resonances must then be found by numerical integration after substituting Eq. (\ref{breitwignercrosssection}) into Eq. (\ref{generalrate}). If transitions to several final states contribute to the total cross section, then the total reaction rate is given by the incoherent sum over the individual transitions, where the rate contribution for each transition can be calculated by integrating Eq. (\ref{generalrate}) numerically. It should be emphasized that Eqs. (\ref{breitwignercrosssection})--(\ref{gammapartwidth}) and (\ref{scalinggam}) apply equally to subthreshold states (that is, when $E_r<0$). 
%
\subsection{Interferences}\label{}
When two broad resonances of the same spin and parity are located sufficiently close to each other, their amplitudes may interfere. The total cross section contribution of the two resonances can be estimated by using a simplified two-level dispersion formula, given by \cite{Lan58}
\begin{equation}
\sigma(E) = \sigma_1(E) + \sigma_2(E) \pm 2 \sqrt{\sigma_1(E) \sigma_2(E)} \cos(\delta_1 - \delta_2)  \label{interf}
\end{equation}
where $\sigma_i(E)$ is obtained from Eq.~(\ref{breitwignercrosssection}). The resonance phase shifts can be calculated using
\begin{equation}
\delta_i = \arctan \left[ \frac{\Gamma_i(E)}{2(E-E_{ri})} \right] \label{interphase}
\end{equation}
with $E_{ri}$ and $\Gamma_i$ the resonance energy and total width of resonance $i$ respectively. The reaction rate contribution of the two interfering resonances must be found by numerical integration after substituting Eq. (\ref{interf}) into Eq. (\ref{generalrate}). Note that Eqs.~(\ref{interf}) and (\ref{interphase}) also apply to subthreshold states.
%
\section{Classical reaction rates}\label{classicalrates}
With the formalism provided in the previous section, we will now summarize some established techniques of estimating reactions rates.  Depending on what kind of nuclear physics information is available the details usually vary and, consequently, each reaction has been treated as a special case. Nevertheless we will focus here on the overall picture. The main goal of this section is to emphasize the statistical shortcomings of the established procedures.
%
\subsection{Established procedures}
Usually one starts by compiling the primary nuclear data needed to calculate reaction rates: the Q-value, resonance and excitation energies, level spins and parities, resonance strengths, particle and $\gamma$-ray partial widths, spectroscopic factors, reduced widths, and non-resonant S-factors for direct capture or for broad resonance tails. When measured directly, each nuclear physics property for a given level will have an associated mean value and an uncertainty. Frequently, more than one measurement of the same quantity has been performed so that either a weighted average can be derived, or each series of measurements can be normalized separately to some standard values. When in exceptional cases no experimental information is available and a quantity has to be adopted from theory, it is frequently possible to estimate approximate uncertainties by systematically comparing experimental and theoretical results for nearby levels. The mean values are then used to calculate various contributions to the total reaction rates  according to the expressions given in Sec. \ref{formalism}, that is, narrow and broad observed resonances, unobserved resonances between the particle threshold 
and the lowest-lying observed resonance, subthreshold states, nonresonant processes, and possible interferences between different amplitudes.

Specifically, once a nonresonant S-factor has been expanded according to Eq. (\ref{Sfactorpoly}), the nonresonant rates are calculated from Eqs. (\ref{nonresrateexpr})-(\ref{cutoff}). Measured energies and strengths of narrow resonances can be used directly in Eq. (\ref{narresrateexpr}) to find their reaction rate  contribution. For threshold states (that is, unobserved narrow low-energy resonances) it is usually possible to estimate the resonance strength according to Eqs. (\ref{resstrength}), (\ref{partpartwidth})-(\ref{iliadisexpr}) if the reduced width (or the spectroscopic factor) can be determined by independent means, for example, from transfer reactions. Frequently, only two reaction channels are energetically allowed such that for threshold levels the particle partial width, $\Gamma_p$, is much smaller than the $\gamma$-ray partial width, $\Gamma_{\gamma}$. Thus we find $\omega \gamma \approx \omega \Gamma_p$ and the reaction rate contribution is given by Eq. (\ref{narresrateexpr}). Similarly, the S-factor of observed broad resonances can be integrated numerically according to Eq. (\ref{breitwignercrosssection}), where the partial widths are deduced from the measured resonance strength, $\omega \gamma$, and the total resonance width, $\Gamma$. The contribution of unobserved broad resonances or subthreshold states can be estimated from Eq. (\ref{breitwignercrosssection}) once particle reduced widths (from transfer studies or elastic scattering) and $\gamma$-ray partial widths (from $\gamma$-ray decay studies or measurement of mean lifetimes, $\tau = \hslash/\Gamma$) are known.
Sometimes the required nuclear properties are not known for the levels of astrophysical interest, which is frequently the case in reactions involving short-lived targets. In such cases the necessary information may be adopted from corresponding levels in the mirror or analog nuclei (see, for example, Iliadis et al. \cite{Ili99} for more information on this procdure).

We have addressed so far only the calculation of what is called in the literature {\it recommended} reaction rates. The estimation of {\it reaction rate uncertainties} from uncertainties in the nuclear physics input values (resonance energies and strengths, S-factors, spectroscopic factors, and so on) is not as straightforward. In fact, no generally accepted procedure exists, which certainly reflects a number of major underlying problems. We would like to point out that no reaction rate uncertainties are given at all in Caughlan and Fowler \cite{CF88}, while no specific information is provided in the NACRE work \cite{Ang99} on how the ``upper" and ``lower" limits have been derived from uncertainties in the nuclear physics input.
%
\subsection{Conventional meaning of ``upper" and ``lower" reaction rate limits \label{conmean}}
Let us start by considering a simple, although unrealistic, situation. At some temperature the total reaction rate is given by the contribution of a single narrow resonance. Furthermore, suppose that the uncertainty in the resonance energy, $\Delta E_r$, is negligible compared to the uncertainty in the measured resonance strength, $\Delta \omega \gamma$, where the latter uncertainty typically amounts to $\pm$10-25\%. The resonance strength enters linearly in Eq. (\ref{narresrateexpr}) and thus the uncertainty in the total rate is given by the uncertainty in the resonance strength. However, in general the uncertainty in the resonance energy cannot be disregarded. The energy enters exponentially in Eq. (\ref{narresrateexpr}) and the uncertainty in the total rate can then be found from $\Delta E_r$ and $\Delta \omega \gamma$ by using standard analytical uncertainty propagation techniques. In almost all cases of practical interest, however, many different reaction rate contributions must be taken into account when calculating the rate uncertainty. Recall that some of these rate contributions can only be found from numerical integration, as outlined in Sec. \ref{formalism}. Consequently, in reality the situation becomes sufficiently complex that analytical uncertainty propagation methods are almost never applied in practice (see Sec. \ref{analyticrates} for an exception). Instead, a frequently applied procedure is to find intuitively the major sources of rate uncertainties and to vary these input parameters individually in order to guess some kind of boundaries for the resulting reaction rate, which are then referred to as ``upper" and ``lower" rate limits. 

Clearly, the procedure just described lacks a rigorous statistical meaning. What precisely do these rate limits quantify? Do they reflect properties of the probability density function associated with the total reaction rate? And what is the corresponding confidence level or coverage probability? For most reaction rates published to date these questions have no clear answers. For example, if the published value of a ``lower" reaction rate limit amounts to $2\times10^{-5}$ cm$^3$~mol$^{-1}$~s$^{-1}$, does this mean that the rate cannot be smaller than this value? It should be obvious to the reader that there is no sharp cutoff in the above example and that the rate can indeed become smaller than the ``lower" limit, since the probability density function of the total reaction rate is determined by a continuous probability density function for each measured or estimated nuclear physics input parameter. 

Until recently, stellar modelers were only interested in the recommended reaction rate and reaction rate uncertainties were disregarded entirely. This has changed, especially over the past decade, and recently more emphasis is placed on investigating the influence of reaction rate uncertainties on stellar energy production and nucleosynthesis (for example, see Refs. \cite{The98,Hof99,Arn99,Jor03,Sas05,Her06}). Nevertheless, the published ``upper" and ``lower" limit values of the total reaction rate are interpreted by the astrophysics community as sharp boundaries. For example, a study of globular cluster ages using Monte Carlo techniques \cite{Kra03} sampled over a uniform (that is, a constant) probability density function (Sec. \ref{otherdist}) between the published ``lower" and ``upper" rate limit of the $^{14}$N(p,$\gamma$)$^{15}$O reaction (with zero probabilities outside these boundaries). Similarly, recent investigations of standard big bang nucleosynthesis \cite{Coc02} or of the influence of proton capture reaction rate uncertainties on the hot bottom burning in intermediate-mass AGB stars \cite{Izz07} assumed a uniform probability density function between upper and lower reaction rate limits. We do not argue that this procedure is wrong since there is no obvious alternative considering that only the lower limit, recommended value and upper limit of the total reaction rate are usually presented in the literature. However, we argue here that the published information is incomplete and that it is of crucial importance to provide the complete probability density function of the total reaction rate at each value of the temperature.
%
\subsection{Analytical reaction rate uncertainties} \label{analyticrates}
The reaction rate uncertainty analysis in the evaluation of Iliadis et al. \cite{Ili01} was partially based on the analytical method developed by Thompson and Iliadis \cite{Tho99}. This method represented the first extensive step towards a statistically meaningful reaction rate uncertainty estimation. An analytical approach has the advantage that it provides insight with respect to which individual rate contributions precisely dominate the total reaction rate. It also allows for studying the correlation between different uncertainties. The formalism of Thompson and Iliadis \cite{Tho99} was incorporated into a computer code, \texttt{RateErrors}, which was made available to the nuclear astrophysics community. The reader is referred to Ref. \cite{Tho99} for details. Here we will focus on a number of important issues.

It may be obvious to the reader that an analytical uncertainty propagation is not straightforward considering the complexity of the expressions provided in Sec. \ref{formalism}. Thus Thompson and Iliadis \cite{Tho99} were required to apply certain approximations and assumptions in their derivations in order to keep the uncertainty propagation tractable. It was demonstrated in Ref. \cite{Tho99} that the formalism works well for narrow resonances when the uncertainty in the resonance energy is relatively small (say, a few keV). Another interesting aspect of this work was their assumption (without proof) that the probability density function of the resulting total reaction rate is given by an expression of the form
\begin{equation}
f(x) = f(x_m)~ e^{-[x_m \ln(x/x_m)]^2/(2\sigma^2)},~~~~~x>0    \label{quasilog}
\end{equation}
where $x_m$ and $\sigma$ are the most probable value and the standard deviation, respectively, of the total reaction rate. Note that Eq.~(\ref {quasilog}) is symmetric on a logarithmic scale and asymmetric on a linear scale. Once an expression for the probability density function is adopted it is a simple matter to estimate the confidence level, that is, the cumulative probability between the uncertainty boundaries.

Over the past few years, experience with the code \texttt{RateErrors} has clearly shown its limitations and shortcomings. Although the code had been extended to include nonresonant reaction rate contributions, it does not, by construction, allow for numerically integrated rate contributions. Neither does it work when the uncertainty in the resonance energy becomes relatively large in which case the  first-derivative approximations of Ref. \cite{Tho99} break down. For this reason we felt compelled to develop a new formalism that applies more generally and is not subject to the restrictions discussed above.
%
\subsection{Problem of ``upper limits" in nuclear physics input} \label{ulproblem}
The problems referred to above are significantly exacerbated by a question we have not addressed so far, that is, how to interpret and to incorporate measured ``upper limits" of nuclear physics quantities into the reaction rate formalism. This problem is most important for expected, but yet unobserved, resonances at low energies (that is, for levels near the particle threshold) where direct cross section measurements are difficult. 

Suppose an ``upper limit" value has been determined for a nuclear property, such as a resonance strength or a spectroscopic factor. How does this upper limit enter in the uncertainty propagation in order to estimate the total reaction rates? The strategy adopted by the NACRE collaboration \cite{Ang99} was the following. First, the total reaction rate is calculated according to the formalism given in Sec. \ref{formalism} by excluding all ``upper limits" of input quantities, that is, the rate contributions from such resonances is set equal to zero; this provides the ``lower limit" of the total rate. Second, the ``upper limit" of the total reaction rate is found by including all the upper limits of input quantities. Third, a recommended reaction rate is found by including all contributions considered under the second step, except that all upper limit contributions of expected resonances are multiplied by an (arbitrary) factor of $0.1$. A similar procedure had been adopted in the evaluation of Iliadis et al. \cite{Ili01}. 

The analytical uncertainty analysis method of Thompson and Iliadis \cite{Tho99} does not allow for the uncertainty propagation of  ``upper limits" in input quantities. The problem really consists of how to interpret and to use a measured upper limit. For example, what does it mean if the literature reports an upper limit of $C^2S<0.05$ for the spectroscopic factor? Does it mean that larger values are excluded? This is certainly not the case but this is precisely how the reported values are being interpreted so far in nuclear astrophysics.

Let us be more precise in our observations. In the overwhelming number of cases, the value of an ``upper limit" for a nuclear physics quantity is presented in the literature without further information. Clearly, the reported number by itself has no rigorous statistical meaning. A piece of information that is obviously missing is the confidence level associated with this upper limit value. In a few selected cases, a value for the confidence level is indeed provided in the literature. Even if this is the case, the most important piece of information is still missing, that is, the probability density function used to derive both the ``upper limit" and the associated confidence level. We strongly urge the community to consider in the future this issue carefully and to present the complete information when reporting on a null-result: (i) the value of the upper limit, (ii) the associated value of the confidence level, and (iii) the probability density function used for deriving the values referred to under (i) and (ii). 

An interesting observation in this regard was reported by Thompson and Iliadis \cite{Tho99}. They pointed out that the statistical distribution of reduced widths, $\gamma^2$, and spectroscopic factors, $C^2S$, are known to be described by a Porter-Thomas distribution (see below for details) and that this circumstance could be used in order to draw a random sample of $\gamma^2$ or $C^2S$ in the absence of any other information on the properties of the level in question. This idea was not pursued 
in Ref. \cite{Tho99} but, in fact, represents a starting point for the statistical treatment
of upper limits in the present work.
%
\section{Statistical distributions}\label{statdistr}
Having described in the previous section some established procedures of reaction rate estimation, we will now turn the attention to our method of calculating reaction rates. We will begin with a brief discussion of statistical distributions. Although this information is provided in many books on statistics (for example, see Ref. \cite{Cow98}), it is worthwhile to summarize here the important expressions because they will be referred to in the following discussion and in Papers II and III. 

The expectation value (or {\it mean}) for any probability density function $f(x)$ and the corresponding variance (or square of the {\it standard deviation}) are given by
\begin{equation}
E[x] = \int_{-\infty}^{+\infty} x f(x) dx,~~~~~~
V[x] = \int_{-\infty}^{+\infty} (x-E[x])^2 f(x) dx \label{meanvar}
\end{equation}
where $f(x)$ is normalized to unity over the entire sample space. The {\it cumulative distribution} related to $f(x)$ is
\begin{equation}
F(x) = \int_{-\infty}^{x} f(x^{\prime}) dx^{\prime} \label{cumdistr}
\end{equation}
and corresponds to the probability for the random variable to assume a value less than or equal to $x$. The {\it quantile} of order q, $x_q$, is defined as the value of the random variable $x$ such that $F(x_q)=q$, with $0\leq q \le1$. In other words, the quantile is equal to the inverse function of the cumulative distribution, $x_q = F^{-1}(q)$. A frequently used quantile is $x_{1/2}$, called the {\it median} of $x$, which corresponds to observing $x$ with equal probabilities below or above $x_{1/2}$. 
%
\subsection{Gaussian distribution}\label{gaussian}
The Gaussian (or normal) probability density function of a continuous random variable $x$ is defined by
\begin{equation}
f(x) = \frac{1}{\sigma \sqrt{2\pi}} e^{-(x-\mu)^2/(2\sigma^2)}  \label{gaussianpdf}
\end{equation}
and has two parameters, $\mu$ and $\sigma$. The first parameter determines the location of the distribution maximum and is found to be equal to the expectation value, while the second parameter controls the distribution width and can be shown to be equal to the square-root of the variance. Thus
\begin{equation}
E[x] = \mu,~~~~~
V[x] = \sigma^2 \label{Gausimple}
\end{equation}
The cumulative distribution of the Gaussian cannot be presented in analytical form, and must be computed numerically by using the error function \cite{erf}. For example, from standard tabulations one finds for the percentage probability that a point is located within $1\sigma$ of the mean (that is, $\mu - \sigma < x < \mu + \sigma $) a value of 68.3\%, while one finds a value of 95\% within $2\sigma$ of the mean. 

The Gaussian distribution is the most frequently used probability density function. Part of its appeal arises from its simplicity (that is, symmetry and bell-shape), leading to straightforward visualizations. From a more rigorous statistical point of view, its importance originates from the {\it central limit theorem}. The theorem states that the sum of $n$ independent continuous random variables $x_i$ with means $\mu_i$ and standard deviations $\sigma_i$ becomes a Gaussian random variable in the limit of $n \rightarrow \infty$, independent of the form of the individual probability density functions of the $x_i$. Many measurement uncertainties are treated as Gaussian random variables if it can be assumed that the total uncertainty is given by the sum of a large number of small contributions. For example, it is reasonable to assume that a measured resonance energy is Gaussian distributed (see Sec. 5.1.1.).

An obvious, but sometimes overlooked, property of the Gaussian distribution is that Eq. (\ref{gaussianpdf}) is defined for $ - \infty < x < \infty $. Thus there is a finite probability that a point is located in the negative region. This issue becomes important when describing physical quantities with a Gaussian density distribution. Since, for example, a negative resonance strength or partial width is obviously unphysical, the Gaussian probability density function is sometimes truncated at the origin (that is, it is set equal to zero for $x < 0$). However, such a procedure is highly suspicious since it clearly introduces a bias by changing the values of $E[x]$ and $V[x]$.
%
\subsection{Lognormal distribution}\label{lognormal}
Suppose that a continuous random variable, given by $y = \ln(x) $, is Gaussian distributed. The variable $x$ will then follow the lognormal distribution, given by
\begin{equation}
f(x) = \frac{1}{\sigma \sqrt{2\pi}} \frac{1}{x} e^{-(\ln x - \mu)^2/(2\sigma^2)}  \label{lognormalpdf}
\end{equation}
The lognormal distribution is defined by the two parameters $\mu$ and $\sigma$. The first determines the location of the distribution, while the second controls the width. Note, that the parameters $\mu$ and $\sigma$ do not represent the mean value and standard deviation of the lognormal distribution, but of the corresponding Gaussian distribution for $\ln(x)$. In terms of these two parameters, the expectation value and the variance of the lognormal distribution are given by
\begin{equation}
E[x] = e^{(2\mu + \sigma^2)/2},~~~~~
V[x] = e^{(2\mu + \sigma^2)}~\left[e^{\sigma^2} - 1 \right] \label{logon}
\end{equation}
Equivalently, the values of $\mu$ and $\sigma$ can be found from $E[x]$ and $V[x]$ by using
\begin{equation}
\mu = \ln(E[x])-\frac{1}{2}\ln \left(1+\frac{V[x]}{E[x]^2} \right),~~~~
\sigma = \sqrt{\ln \left( 1 + \frac{V[x]}{E[x]^2} \right)} \label{lognormpar}
\end{equation}
Since the central limit theorem predicts a Gaussian probability density for a random variable if it is given by the {\it sum} of a large number of small contributions, it follows directly  that a random variable will be distributed according to a lognormal density function if it is given by the {\it product} of many factors. For example, consider the estimation of a resonance strength from the measured thick-target yield. The strength is given by the products and quotients of the following positive quantities: the measured number of counts, the integrated beam charge, a detector efficiency, a stopping power, and so on. If the random uncertainties of these quantities are independent, it is reasonable to assume a lognormal probability density function for the derived resonance strength.

A few more comments will be helpful for the discussions in Sec. \ref{MonteCarlo} and in Paper II. (i) The lognormal density function is defined only in the range of $ 0 \leq x < \infty $ and thus has the desirable property of describing physical quantities that are manifestly positive (for example, a resonance strength).  (ii) For a sample of data, $\left\{ z_1, z_2, ..., z_n \right\} $, that is lognormally distributed, the geometric mean, $\mu_g$, and the geometric standard deviation, $\sigma_g$, are given by
\begin{equation}
\mu_g \equiv \sqrt[n]{z_1 \cdot z_2 \cdot ... \cdot z_n} = e^{\mu},~~~~
\sigma_g \equiv \exp \left[\sqrt{\frac{1}{n} \sum_{i=1}^n \left( \ln \frac{z_i}{\mu_g} \right)^2 }\right] = e^\sigma \label{geomeanex}
\end{equation}
(iii) The {\it median} value of the lognormal density function is given by $e^{\mu}$, while for a coverage probability of 68\% the lower and upper bounds are given by $\mu_g/\sigma_g=e^{\mu-\sigma}$ and $\mu_g\sigma_g=e^{\mu+\sigma}$, respectively. (iv) For a coverage probability of 68\%, the {\it factor uncertainty} with respect to the median (or geometric mean) is given by $f.u.=e^{\mu+\sigma}/e^{\mu}= e^{\mu}/e^{\mu-\sigma}=e^{\sigma}$ or $\sigma=\ln (f.u.)$; for example, factor uncertainties of 2, 10 and 100 correspond to values of $\sigma=0.69$, 2.3 and 4.6, respectively. (v) For two independent, lognormally distributed, random variables $x_1$ and $x_2$, with location and spread parameters of $\mu_1$, $\sigma_1$ and $\mu_2$, $\sigma_2$, the product $\alpha x_1^{\beta_1}x_2^{\beta_2}$ (where $\alpha>0$) is also lognormally distributed, with location and spread parameters of $\mu=\ln \alpha + \beta_1\mu_1+\beta_2\mu_2$ and $\sigma^2=\beta_1^2\sigma_1^2+\beta_2^2\sigma_2^2$, respectively. The use of lognormal distributions for describing factor uncertainties is discussed in App. B of Iliadis et al. \cite{Ili99}. For more information on the lognormal distribution, see Ref. \cite{Cro88}.

\begin{figure}[]
\includegraphics[height=13.5cm,angle=-90]{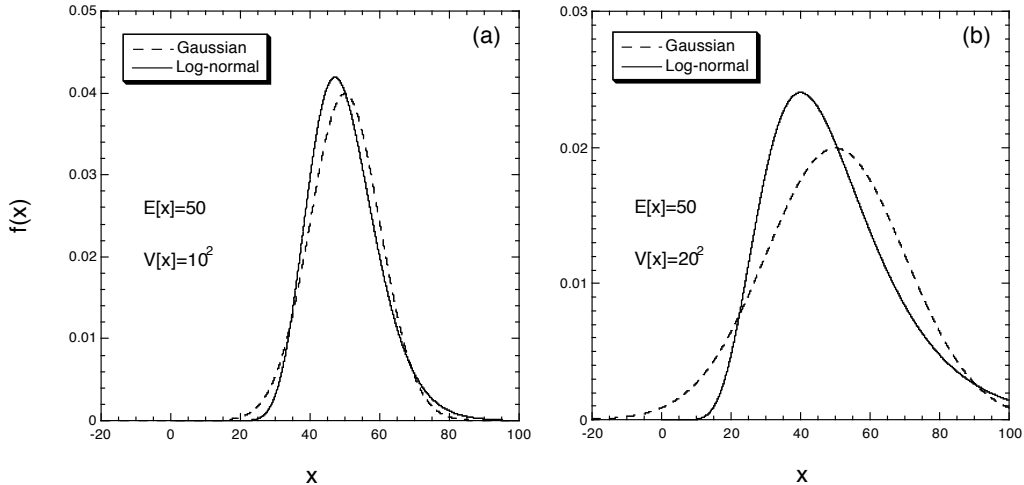}
\caption{\label{fig:GauLog} 
Comparison of Gaussian (dashed lines) and lognormal (solid lines) probability density functions; (a) $E[x]=50$ and $V[x]=10^2$, and (b) $E[x]=50$ and $V[x]=20^2$. See text.}
\end{figure}

The lognormal distribution is skewed, that is, it is asymmetric. This may be the reason for the widespread use of Gaussian distributions even if the data sample is on statistical grounds more properly described by a lognormal distribution. An impression can be obtained from Fig.~\ref{fig:GauLog} which compares Gaussian and lognormal distributions of the same expectation value and variance. In part (a) the values chosen are $E[x]=50$ and $V[x]=10^2$, that is, the standard deviation amounts to 20\% of the mean value. For the Gaussian distribution these values are equal to $\mu$ and $\sigma^2$, respectively, according to Eq. (\ref{Gausimple}). For the lognormal distribution, one finds from Eq. (\ref{lognormpar}) the values $\mu=3.8924$ and $\sigma=0.198$. It is obvious that both distributions have very similar shapes. A smaller value of the variance, so that $\sigma \leq 0.1$ for the lognormal distribution, results in two curves that are visually indistinguishable on this scale. Part (b) represents the situation when the variance is relatively large compared to the expectation value. In this case we have $E[x]=50$ and $V[x]=20^2$, that is, the standard deviation amounts now to 40\% of the mean value. We obtain $\mu=3.8378$ and $\sigma=0.385$ for the lognormal distribution, which is clearly skewed. Furthermore, for a coverage probability of 68\% the factor uncertainty amounts here to $f.u.=e^{\sigma}=1.47$. Note that the Gaussian probability density function continues in the negative region, while the lognormal distribution is only defined for positive values of $x$.
%
\subsection{Chi-squared distribution}\label{chisquarepdf}
Consider $k$ independent Gaussian distributed random variables with mean of $\mu_i = 0$ and variance of $\sigma_i^2 = 1$. The sum of their squares is distributed according to a chi-squared distribution, where the parameter $k$ is called the number of degrees of freedom. The chi-squared distribution plays an important role in data fitting in connection with the method of least squares.

\begin{figure}[]
\includegraphics[height=13.5cm,angle=-90]{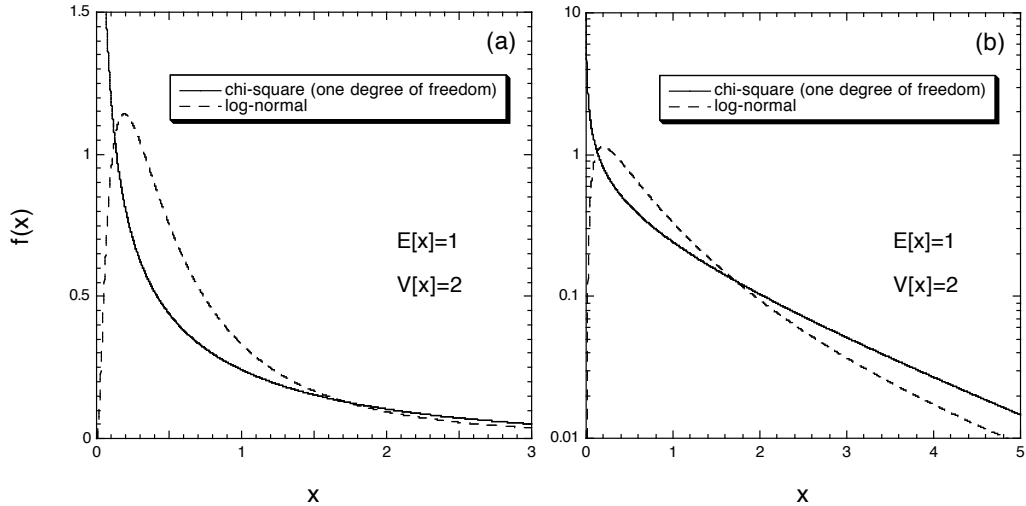}
\caption{\label{fig:chisq} 
Chi-square distribution of one degree of freedom (solid line) on (a) a linear vertical axis scale, and (b) a logarithmic vertical axis scale. For comparison, a lognormal distribution (dashed line) of same expectation value and variance as the chi-squared distribution is shown. The lognormal parameters amount to $\mu=-0.54931$ and $\sigma=1.0482$, according to Eq. (\ref{lognormpar}).}
\end{figure}

In the present work, it will be used in an entirely different context. We are mostly interested in the simplest case: a chi-squared distribution with one degree of freedom ($k=1$), that is, the sum of squares consists of a single term only. In physics this special distribution is also referred to as Porter-Thomas distribution (Sec. \ref{inputdistr}). The chi-squared distribution with one degree of freedom is given by
\begin{equation}
f(x) = \frac{1}{\sqrt{2\pi x}}e^{-x/2} \label{ptdist}
\end{equation}
This distribution is defined over the range of $0 < x < \infty $ and has no adjustable parameters. The expectation value and the variance are given by
\begin{equation}
 E[x]=k=1, ~~~~~~
 V[x]=2k=2
\end{equation}
The chi-squared distribution with one degree of freedom is displayed in Fig. \ref{fig:chisq}. Note that it is normalizable like any other probability density function although it displays a pole at zero.
%
\subsection{Other distributions}\label{otherdist}
A number of other distributions will be discussed briefly in this work. They are either not included in the formalism presented in the following section or are only used here in exceptional circumstances.

The uniform distribution has been mentioned in Sec. \ref{conmean} in connection with previous interpretations of reaction rate uncertainties. The probability density function for the continuous variable $x$ is given by
\begin{equation}
f(x) = 
\begin{cases}
\frac{1}{b-a} & \text{for $a \leq x \leq b$} \\
~~~0            & \text{for $x < a $, $x > b$} \\
\end{cases}
\end{equation}
representing a square function with $x$ uniformly distributed between $a$ and $b$. The expectation value and variance are given by
\begin{equation}
E[x] = \frac{1}{2}(a+b),~~~~~~~
V[x] = \frac{1}{12}(b-a)^2
\end{equation}
The Poisson distribution is a discrete distribution that is derived from the binomial distribution in the limit that the number of independent trials, $N$, is large and the (constant) probability, $p$, for success of each trial is very small so that the product $Np$ remains equal to some finite and constant value $k$ \cite{Hai67}. The probability density function for the discrete variable $n=0, 1, 2, ...$ is then given by
\begin{equation}
f(n) = \frac{k^n}{n!}e^{-k} \label{poiss}
\end{equation}
The expectation value and variance are 
\begin{equation}
E[n] = k,~~~~~~~
V[n] = k
\end{equation}
Important areas of application for the Poisson distribution are nuclear counting experiments, such as radioactive decay and cross section measurements. For example, for many precisely measured half-lives the final uncertainty is dominated by counting statistics. In such cases, the probability density function of the half-life is likely given by a difference of two Poisson distributions (for the total and background count rate). The Poisson probability density function can be approximated by a (continuous) Gaussian distribution if the value of the parameter $k$ is not too small (say, $k>5$). The Poisson probability density, although mentioned in Sec. \ref{ulwgsec}, is not used in the present work.

Finally, the binary probability density function can only assume two discrete values with 
a probability of $1/2$ for each value. It is employed in the present work to describe interfering amplitudes when the sign of the interference term is unknown.
%
\section{Monte Carlo method}\label{MonteCarlo}
We will now discuss a method, based on the Monte Carlo technique\footnote{During the near completion stage of the present work, the effort of Ref. \cite{Smi02} has been brought to our attention. These authors explore a Monte Carlo technique for calculating the rates of the $^{31}$P(p,$\gamma$)$^{32}$S reaction. They address the simplest possible input (that is, the total rate is given as an analytical function of resonance energies and strengths only) and assume lognormal distributions both for the nuclear input and rate output probability density functions. We would like to emphasize that unlike the present work their paper does not consider correlations in the random sampling, numerical rate integration, upper limits of nuclear quantities, interferences between resonances, nonresonant and resonance tail contributions, or deviations of the rate probability density from lognormality; furthermore, while we adopt a normal probability density function for resonance energies, which accounts naturally for subthreshold states (Sec. \ref{resinfosec}), the authors of Ref. \cite{Smi02} assume a lognormal distribution.}, of estimating thermonuclear reaction rates. First, we turn our attention to associating each nuclear property that enters in the estimation of reaction rates (Sec. \ref{formalism}) with a specific probability density function (Sec. \ref{statdistr}). Second, the elusive problem of upper limits for resonance strengths and partial widths is addressed. Third, the Monte Carlo sampling procedure is explained in detail. Finally, some properties for the output distribution of reaction rates will be discussed.

It must be kept in mind that our approach of estimating reaction rates requires (as any other method would) a certain minimum amount of nuclear physics input. For example, if information on crucial resonance parameters or interfering amplitudes is entirely missing, then the uncertainties obtained with the present method would not be very meaningful compared to the {\it errors} caused by (unknown) systematic effects. Specific reactions that had to be excluded from the present evaluation will be discussed in Sec. 3 of Paper II, so that the reader can get an impression on their scope. 
\subsection{Input statistical distributions of nuclear physics data}\label{inputdistr}
\subsubsection{Resonance energies}\label{resinfosec}
Almost all of the resonance energies used in the present work are obtained from (i) the 50\% point on the low-energy edge of the measured thick-target yield curve, or (ii) the measured excitation energy of the corresponding compound state by using $E_r = E_x - Q$. In the first case, the probability density function of the resonance energy in a given measurement is most likely described by a Gaussian distribution since it can be assumed that the total uncertainty is given by the sum of several small contributions (Sec. \ref{gaussian}), for example, arising from the beam energy calibration, the measured yields, the fitting of the yield curve to find the 50\% point, target inhomogeneities and dead layers, and so on. In the second case, which occurs frequently for low-energy resonances or reactions involving short-lived target nuclei, both the excitation energy and the reaction Q-value can be assumed to be Gaussian distributed, so that the difference is again described by a Gaussian probability density function. Thus we associate the reported or derived values of the resonance energy and the corresponding (``1$\sigma$") uncertainty with the parameters $\mu$ and $\sigma$ of Eq. (\ref{gaussianpdf}), respectively.

We already mentioned in Sec. \ref{gaussian} that a Gaussian distribution predicts a finite probability for obtaining a negative random variable. This does not pose a problem for resonance energies since as soon as a negative value of $E_r$ is sampled in the Monte Carlo method, we treat the resonance as a subthreshold level and switch to the appropriate formalism, as discussed in Sec. \ref{numintrates}. 
%
\subsubsection{Resonance strengths}\label{reswgex}
Resonance strengths have either been measured directly or are estimated from partial widths. In the first case, we renormalized as far as possible all the literature $\omega\gamma$ values according to the standard resonance strengths listed in Tab.~4.12 of Ref.~\cite{Ili07}. Our procedure differs from that of NACRE \cite{Ang99}, where in most cases a weighted average of all reported $\omega\gamma$ values was adopted, regardless of any systematic deviations between the different data sets. Uncertainties of measured resonance strengths range from 4\% for very careful studies to more typical values of 15-20\%. These uncertainties are usually interpreted as $1\sigma$ uncertainties of a Gaussian distribution. However, it is clear that such an interpretation will result a fraction of the time in negative resonance strengths and, consequently, in negative reaction rates according to Eq. (\ref{narresrateexpr}). It is worthwhile to recall how a resonance strength is estimated in an experiment. Its value is given by {\it products} or {\it quotients} of positive and independent quantities, such as a measured number of counts, a stopping power, a detection efficiency, an integrated beam charge, and so on. As already mentioned in Sec. \ref{lognormal}, the distribution of resonance strengths will then tend towards a lognormal probability density function rather than a Gaussian. Thus we associate the measured resonance strength and its corresponding uncertainty with the expectation value and the square root of the variance, respectively, of a lognormal distribution. The corresponding parameters $\mu$ and $\sigma$ are then obtained from Eq. (\ref{lognormpar}). Recall that for relatively small variances a lognormal distribution is almost indistinguishable from a Gaussian, but the former probability density function has the desired property of predicting only positive values for the resonance strength (Sec. \ref{lognormal}). 

If a resonance has not been observed yet, which occurs frequently for low-energy resonances or reactions involving short-lived target nuclei, then its strength can be estimated from the partial widths by using Eqs. (\ref{resstrength}), (\ref{partpartwidth})--(\ref{gammapartwidth}). The estimate of the particle partial width requires knowledge of the reduced width or the spectroscopic factor\footnote{The reduced width and the spectroscopic factor are also related to the asymptotic normalization coefficient (ANC). The relationship of reduced width and ANC is given by Eqs. (55) and (60) of Ref. \cite{Muk99}, while the relationship of spectroscopic factor and ANC can be obtained from $(ANC) = (C^2S)^{1/2}~(ANC)_{sp}$, where $(ANC)_{sp}$ denotes the single-particle ANC \cite{Muk01}.}, which can be measured in direct (transfer) reaction studies. Similarly, the $\gamma$-ray partial width can be estimated from the reduced transition probability, which may be obtained from $\gamma$-ray decay studies. If the level properties for the states of astrophysical interest have not been measured yet, then it is frequently possible to adopt the required values of the spectroscopic factor and the reduced transition probability from the corresponding levels in the mirror nuclei. More information on this method can be found in Iliadis et al. \cite{Ili99}. For the probability density functions of the particle and $\gamma$-ray partial widths we assume again lognormal distributions for reasons similar to those given in connection with measured resonance strengths. The derived values of $\Gamma_{\lambda c}$ or $\Gamma_{\gamma}$ represent expectation values. For the square root of the variance we assume a value of 40\% for the particle partial width and 50\% for the $\gamma$-ray partial width. The choice of these values is supported by a systematic comparison of partial widths \cite{Ili99,Hal04} and by an uncertainty analysis of measured spectroscopic factors \cite{Tho99}. The parameters $\mu$ and $\sigma$ of the lognormal distribution are then again found from Eq. (\ref{lognormpar}). In some cases we have to resort to shell model calculations of spectroscopic factors or reduced transition probabilities. For the sake of consistency, values of 40\% and 50\% are used for the square root of the variance of $\Gamma_{\lambda c}^{SM}$ and $\Gamma_{\gamma}^{SM}$, respectively.

In exceptional cases, a ``best" value can be derived for a partial width, but the associated uncertainty can only be estimated within a certain factor. For example, assume that the best value of a partial width amounts to $1.0\times10^{-5}$ eV and that this value is uncertain by a factor of 3. For a lognormal distribution, the factor uncertainty relates to the {\it median} value rather than the expectation value, as explained in Sec. \ref{lognormal}. Thus we can assume that the interval between $\frac{1}{3}(1.0\times10^{-5})=3.3\times10^{-6}$ eV and $3(1.0\times10^{-5})=3.0\times10^{-5}$ eV contains a coverage probability of 68\%. Interpreting $1.0\times10^{-5}$ eV as the median value and $3$ as the factor uncertainty, we find for the lognormal parameters $\mu=\ln(1.0\times10^{-5})=-11.513$ and $\sigma=\ln(3)=1.0986$. According to Eq. \ref{logon}, this yields $E[x]=1.8\times10^{-5}$ eV and $\sqrt{V[x]}=E[x]\sqrt{e^{{\sigma}^2}-1}=2.8\times10^{-5}$ eV. The square root of the variance is larger than the expectation value, indicating that the lognormal distribution is highly skewed.
%
\subsubsection{Nonresonant S-factors}
The nonresonant reaction rate formalism of Sec. \ref{nonresrates} can be used to calculate the contribution of direct nuclear processes. The most important nonresonant process in the present context is called direct radiative capture. This relatively weak process has been observed only in a few reactions (see Ref. \cite{Ili04} and references therein). In the overwhelming number of cases, the direct capture cross section must be estimated from experimental nuclear structure information. The cross section is given by
\begin{equation}
\sigma_{total}^{DC} = \sum_{j} \sum_{\ell_{i} \ell_{f}} 
               C^{2}S_{j}(\ell_{f})~\sigma^{DC}_{calc,j}(\ell_{i},\ell_{f})
\end{equation}
where the incoherent sum is over all orbital angular momenta $\ell_{i}$ and $\ell_{f}$ of the initial scattering state and the final bound state, respectively, and over all final bound states $j$; $C^2S_j$ is the experimental spectroscopic factor for state $j$ and $\sigma_{calc,j}^{DC}$ is the cross section for a specific transition calculated by using a single-particle potential model. The reaction rate contribution can then be found from numerically integrating Eq. (\ref{generalrate}) after substitution of the total direct capture cross section or, equivalently, from Eqs. (\ref{Sfactorpoly})-(\ref{Sfactorterm}) after converting the cross section to an S-factor. We employed the latter procedure and assumed a lognormal probability density function for the total S-factor in Eq. (\ref{Sfactorpoly}). For the square root of the variance we adopt a value of 40\%, based on a systematic comparison of experimental spectroscopic factors from direct capture and from transfer reaction studies \cite{Ili04}. A value of 40\% for the square root of the variance of the S-factor is also used in exceptional cases where $C^2S_j$ has to be extracted from the shell model.
%
\subsection{Upper limits}
We will now return to the elusive problem of how to include in the reaction rate uncertainty analysis an upper limit for a resonance strength or a partial width. Two situations are of practical interest. First, a resonance is not observed in a search and all that can be obtained from the experimental spectrum of emitted particles or $\gamma$-rays is the total number of background counts in the region of interest. Second, not only is a resonance not observed in a direct search, but the corresponding compound level is not even populated in a transfer reaction study. Again, all that is obtained from the transfer experiment is the total number of background counts in the region of interest. The latter case is by far the most important one and will be addressed first. 

\subsubsection{Upper limits of partial widths}

Assume first that absolutely no experimental information is available on a particle or $\gamma$-ray partial width. All that is known is that a nuclear level occurs at an energy that may or may not strongly influence the total reaction rates. One may be tempted to describe the probability density function for this situation by using a uniform distribution (Sec. \ref{otherdist}), implying a constant probability between zero and some (perhaps dictated by theory) upper limit of $b$. This choice would imply a mean value of $b/2$ and an equal probability for obtaining values below and above $b/2$. However, such an assumption contradicts the predictions of nuclear statistical models, as will be explained in the following.

According to Eqs. (\ref{partpartwidth}) and (\ref{gammapartwidth}) the particle and $\gamma$-ray partial widths are determined by the dimensionless reduced width $\theta^2$ (or the spectroscopic factor $C^2S$) and the reduced transition probability $B$, respectively. Either of these quantities represents a square of an amplitude that is proportional to a matrix element of the nuclear Hamiltonian. The matrix element is equal to an integral over the nuclear configuration space. If the wave functions are sufficiently complex the matrix element will have contributions from many different parts of the configuration space, with the sign and magnitude of a particular contribution being random from level to level and independent in sign and magnitude from all other parts. According to the central limit theorem (Sec. \ref{gaussian}) the probability density function of the nuclear matrix element will be approximately Gaussian with an expectation value of zero. Therefore, it follows immediately that the probability density function for $\theta^2$ or $B$, that is, the square of the amplitude, is given by a chi-squared distribution with one degree of freedom (Sec. \ref{chisquarepdf}). Furthermore, the variance of the Gaussian amplitude distribution is just the {\it local mean value} of the reduced width, $\langle \theta^2 \rangle$, or transition probability, $\langle B \rangle$, for a given single channel (since $V[x]=E[x^2]-E[x]^2$ and $E[x]=0$ for a Gaussian centered at zero). 

These arguments were first presented in Ref. \cite{Por56} and the probability density function is also known as {\it Porter-Thomas distribution}. It is given by Eq. (\ref{ptdist}) where the random variable is equal to the ratio of reduced width or transition probability and their respective local mean value. For example, we may express Eq. (\ref{ptdist}) for a particle channel in terms of the variable $\theta^2$ and find
\begin{equation}
f(\theta^2) =  \frac{c}{\sqrt{\theta^2}} e^{-\theta^2 / (2 \langle \theta^2 \rangle)} \label{renormPT}
\end{equation}
with $c$ denoting a normalization constant. The distribution implies that the reduced widths for a single reaction channel, that is, for a given nucleus and set of quantum numbers, vary by several orders of magnitude with a higher probability the smaller the value of the reduced width. The Porter-Thomas distribution emerges naturally from the Gaussian orthogonal ensemble of random matrix theory (see Ref. \cite{Guh98} for more information). 

There are numerous publications that provide experimental support for the arguments given above (see, for example, Refs. \cite{Bro81,Shr89,Met91,Cam94,Shr05}). Experimental tests of the Porter-Thomas distribution face essentially the same problem, that is, to find data sets of nuclear statistical properties that are sufficiently large to make meaningful predictions. Recall the assumptions we made above: a distribution of reduced widths (or transition probabilities) for a given nucleus, given orbital angular momentum and channel spin, and so on, follows a Porter-Thomas probability density function. Clearly, it is experimentally difficult to collect enough data under such restrictive assumptions. Nevertheless, the validity of the Porter-Thomas distribution can be regarded as firmly established.

\begin{figure}[]
\includegraphics[height=12cm]{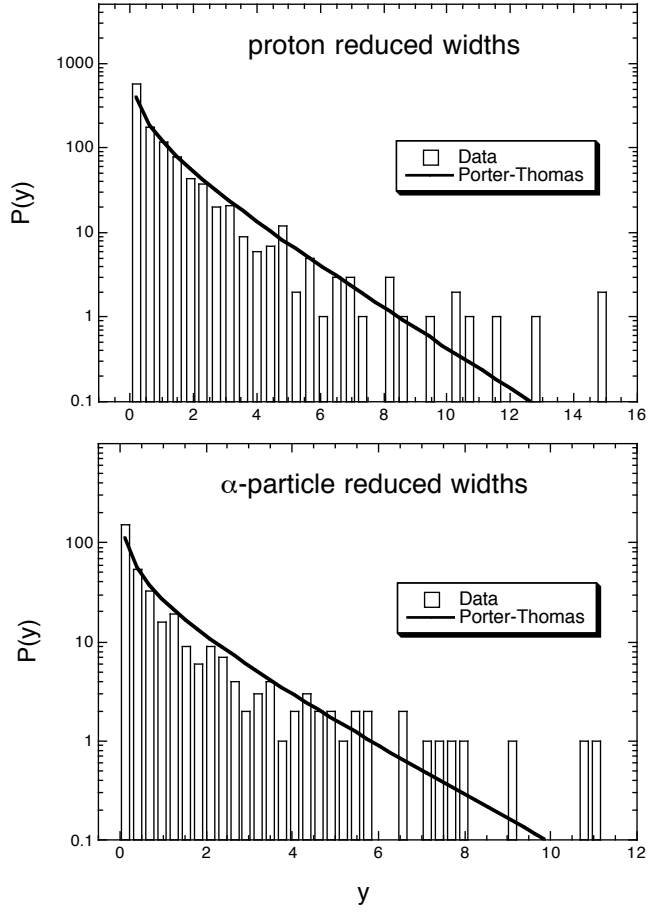}
\caption{\label{fig:theta_renorm} 
Distribution of dimensionless reduced widths for protons (top) and $\alpha$-particles (bottom) of unbound states in $^{24}$Mg, $^{28}$Si, $^{30}$P, $^{32}$S, $^{36}$Ar and $^{40}$Ca. The data are first grouped according to $A$ and $\ell$ and each $\theta^2$ value is then divided by its local mean, that is, $y=\theta^2/\langle \theta^2 \rangle$. The solid curves show Porter-Thomas distributions.}
\end{figure}

Consider, for example, Fig. \ref{fig:theta_renorm} showing the experimental distribution of $1127$ proton and $360$ $\alpha$-particle dimensionless reduced widths of unbound levels in $^{24}$Mg, $^{28}$Si, $^{30}$P, $^{32}$S, $^{36}$Ar and $^{40}$Ca. The data were obtained over the past decades at TUNL by G. Mitchell and collaborators (see Ref. \cite{MIT}, and references therein). We grouped the widths according to nucleus and orbital angular momentum ($\ell$-)transfer, then averaged the widths locally for each group, and finally divided each $\theta^2$ value by the corresponding local mean. The resulting experimental distributions of the random variable $y=\theta^2/\langle \theta^2 \rangle$ are displayed as histograms for protons (top) and $\alpha$-particles (bottom). The solid lines represent Porter-Thomas distributions. The agreement with the data is obvious. One has to be careful when interpreting Fig. \ref{fig:theta_renorm} or similar plots published in the literature. It should be noted that the reduced widths follow a Porter-Thomas distribution only if the nuclear matrix elements have contributions from many different parts of the configuration space. This is clearly not the case for low-lying bound states of nearly closed-shell character or $\alpha$-cluster states where the matrix elements may be dominated by a few large contributions. However, such states exhibit frequently large values of $\theta_p^2$ or $\theta_{\alpha}^2$ and are thus likely to be observed in transfer reaction studies. Neither is there a compelling reason why the reduced widths of isospin-nonconserving particle decays should follow a Porter-Thomas distribution. Such (isobaric analog) resonances normally have very small proton widths and thus make minor contributions to the total rate compared to neighboring resonances. In other words, these exceptional cases are usually not of major concern for the issue of upper limits in nuclear astrophysics.

It must be emphasized that what is of interest in the present work is not the ratio $y=\theta^2/\langle \theta^2 \rangle$, but the value of $\theta^2 = y~\langle \theta^2 \rangle$ which enters directly in Eq. (\ref{partpartwidth}). What is usually presented in the literature is the ratio $y$ and, unfortunately, almost no values of the actual means $\langle \theta^2 \rangle$ are reported. Of course, knowledge of $\langle \theta^2 \rangle$ is required in order to select a random sample of $\theta^2$ from a Porter-Thomas distribution (see below). In Fig. \ref{fig:theta_all} we show the same data as in Fig. \ref{fig:theta_renorm}, but in terms of the variable $\theta^2$, for protons (top) and $\alpha$-particles (bottom). The solid curves now represent least-squares fits of Porter-Thomas distributions to the data, with $f(\theta^2)$ given by Eq. (\ref{renormPT}). The agreement is not as good as in Fig. \ref{fig:theta_renorm} because all $\theta^2$-values have been grouped together, regardless of differences in nuclear mass $A$ or $\ell$-value. Thus the histograms represent sums of Porter-Thomas distributions for different combinations of $A$ and $\ell$. Nevertheless, a single Porter-Thomas distribution describes the total distributions rather well for small values of $\theta^2$, that is, for those levels that will most likely escape detection in a transfer reaction measurement.
\begin{figure}[]
\includegraphics[height=12cm]{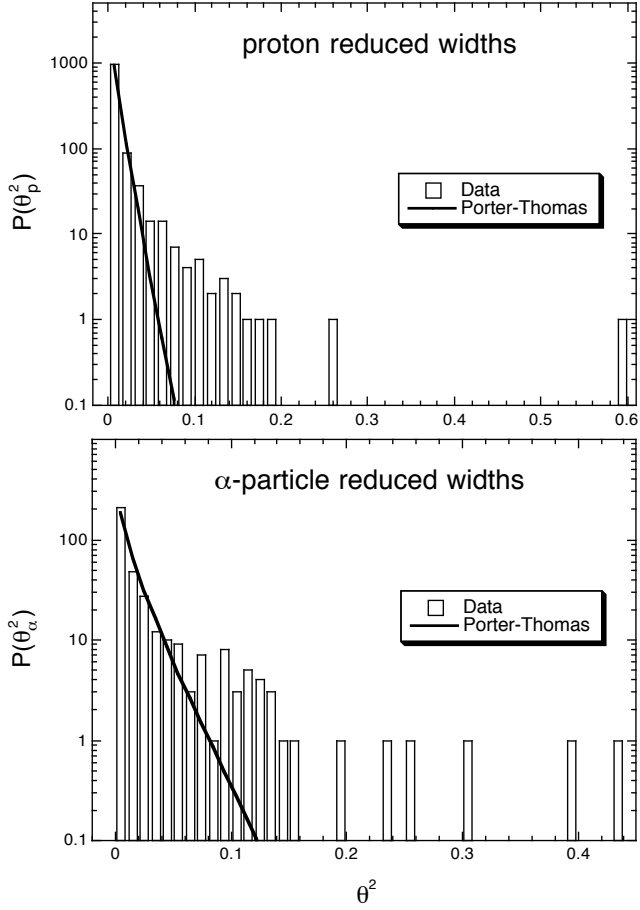}
\caption{\label{fig:theta_all} 
Distribution of dimensionless reduced widths for protons (top) and $\alpha$-particles (bottom) of unbound states in $^{24}$Mg, $^{28}$Si, $^{32}$S, $^{36}$Ar and $^{40}$Ca. All $\theta^2$-values are grouped together, regardless of differences in $A$ or $\ell$. The solid curves represent least-squares fits of Porter-Thomas distributions to the data. The fits result in {\it global} mean values of $\langle \theta^2_p \rangle=0.0045$ and $\langle \theta^2_{\alpha} \rangle=0.010$. These values are adopted for the present reaction rate evaluation.}
\end{figure}
The best-fit mean values of the dimensionless reduced widths for protons and $\alpha$-particles, extracted from the curves displayed in Fig. \ref{fig:theta_all}, amount to $\langle \theta^2_p \rangle=0.0045$ and $\langle \theta^2_{\alpha} \rangle=0.010$, respectively. These values will be adopted in the present reaction rate evaluation. We also performed least-squares fits to individual groups of data, each characterized by a given combination of $A$ and $\ell$. As a result we find indeed some scatter, by a factor of $\approx2-3$, around the values of $\langle \theta^2\rangle$ quoted above. However, at present it is not clear how much of the scatter is caused by inherent differences between the local mean values as opposed to statistical scatter caused by significantly reduced sample sizes. Clearly, we regard the above numerical choices for $\langle \theta^2\rangle$ as a first, preliminary step. More work in this regard is in order\footnote{The large mean value of the proton spectroscopic factor reported in Ref. \cite{Tho99}, $\langle S_p \rangle=0.65$, disagrees with our results and is certainly erroneous. We are now convinced that their procedure of fitting well-known spectroscopic factors of low-lying {\it bound} states in the sd-shell is inappropriate since it is necessarily biased toward values that are too large (that is, levels that are populated {\it strongly} in transfer reactions). There is no obvious reason why such states should be represented by a Gaussian reduced width amplitude distribution according to the arguments given here.}. 

Much of the above discussion focuses on reduced widths although similar arguments apply to the reduced transition probabilities of $\gamma$-ray transitions: for a given nucleus, given $\gamma$-ray multipolarity and transition character (that is, electric or magnetic) the values of $B(\overline{\omega}L)$ are expected to follow a Porter-Thomas distribution. An extra complication is introduced by the fact that the total $\gamma$-ray width is usually given by the sum of $n$ partial $\gamma$-ray widths,
\begin{equation}
\Gamma_{\gamma} = \sum_i^n \Gamma_{\gamma,i}
\end{equation}
Each of the partial widths $\Gamma_{\gamma,i}$ follows a Porter-Thomas distribution, but the local means of $B_i$ may be different depending on the values of $\overline{\omega}$ and $L$. If many partial $\gamma$-ray widths contribute to the total $\gamma$-ray width ($n$ large) one expects, according to the central limit theorem (Sec. \ref{gaussian}), that $\Gamma_{\gamma}$ tends to follow a Gaussian distribution. In principle, it would be straightforward to select a value of $\Gamma_{\gamma}$ at random once the local means $\langle B_i \rangle$ for the individual transitions have been extracted from large experimental data ensembles of reduced transition probabilities. We did not pursue this idea further since precise values of $\Gamma_{\gamma}$ are not crucial for evaluating the reaction rates presented in Paper II. However, estimates of $\langle B_i \rangle$ would clearly be helpful in special circumstances and for future rate evaluations. More studies in this direction may be needed.

We must now relax our initial assumption that no information is available for a given level other than its excitation energy and quantum numbers. Frequently, there exists additional experimental information in the form of an upper limit for a partial width derived, for example, from an upper limit of a spectroscopic factor (or reduced width) measured in a transfer reaction (Sec. \ref{reswgex}). If the probability density function for the upper limit measurement would be known, then one could combine the experimental result with a Porter-Thomas distribution in order to find the overall probability density function. This could be done, for example, by convoluting the two probability densities in question. Of course, as already pointed out in Sec. \ref{ulproblem}, the probability density function on which the upper limit value is presumably based is usually not reported in the nuclear astrophysics literature. Thus we are faced with the problem of how to incorporate in a meaningful way the information available from the literature (that is, a given value of an upper limit) into the reaction rate formalism (Sec. \ref{formalism}). In lack of a more rigorous approach, we simply adopt
\begin{equation}
f(\theta^2) = 
\begin{cases}
\frac{c}{\sqrt{\theta^2}} e^{-\theta^2 / (2 \langle \theta^2 \rangle)} & \text{if $\theta^2\leq\theta^2_{ul}$} \\
0                                             								& \text{if $\theta^2>\theta^2_{ul}$}  \label{PTsampp}
\end{cases}
\end{equation}
That is, we truncate the Porter-Thomas distribution of Eq. (\ref{renormPT}) at the experimental upper limit of the dimensionless reduced width, $\theta^2_{ul}$. The effects of this choice will be discussed in Paper II.

\subsubsection{Upper limits of resonance strengths}\label{ulwgsec}
Suppose one performs a direct search for an expected low-energy resonance, but no noticeable net signal is observed in the spectral region of interest. All that is observed is unwanted background. For each channel in the region of interest the number of counts is distributed according to Poisson statistics (Sec. \ref{otherdist}): if the mean number of counts is $k$ then the probability that a measurement (of signal or background) will give $n$ counts is given by Eq. (\ref{poiss}). In practice, the background is frequently estimated, for example, in a calibration run. It is then subtracted from the total number of observed counts in order to estimate an upper limit on the number of signal counts, that is, $N^{sig}=N^{tot}-N^{bg}$. The rigorous conversion of an upper limit of counts into an upper limit for a resonance strength, $\omega\gamma_{ul}$, should be based on an appropriate probability density function but, as mentioned in Sec. \ref{ulproblem}, the probability density function used to derive a published upper limit value is usually not reported in the literature. For the future we recommend modern procedures of estimating statistically meaningful upper limits (see, for example, Refs. \cite{Fel98,Han99}).

In order to incorporate a previously reported upper limit of a resonance strength into the reaction rate formalism (Sec. \ref{formalism}) we adopted the following procedure: (i) we assume that the upper limit $\omega\gamma_{ul}$ is determined by an upper limit of the (entrance channel) particle partial width, $\Gamma_{a}^{ul}$; (ii) the value of $\Gamma_{a}^{ul}$ is computed from the measured value of $\omega\gamma_{ul}$ by using Eq. (\ref{resstrength}); (iii) the derived value of $\Gamma_{a}^{ul}$ is converted to an upper limit of the dimensionless reduced width, $\theta^2_{ul}$, according to Eq. (\ref{partpartwidth}); (iv) the probability density function of $\theta^2$ is given by a Porter-Thomas distribution, according to Eq. (\ref{PTsampp}). Assumption (i) is usually fulfilled in reactions with only two open channels since direct searches for low-energy resonances result in sufficiently small resonance strength upper limits so that the condition $\Gamma_a \ll \Gamma_{\gamma}$ holds (see Sec. \ref{narresrates}). With the procedure described above the treatment of upper limits for partial widths and resonance strengths is based on the same foundation.
%
%
\subsection{Monte Carlo sampling}\label{MCsampling}
In previous sections we assigned a specific probability density function to each quantity entering in the calculation of the reaction rates (Sec. \ref{formalism}): a Gaussian for resonance energies; a lognormal distribution for measured resonance strengths, nonresonant S-factors and partial widths; a Porter-Thomas distribution for measured upper limits on partial widths; and so on. Once a probability density function is chosen for each input quantity, the total reaction rate and the associated uncertainty can be estimated using standard Monte Carlo techniques (see Ref. \cite{Cow98}). In particular, a random value is generated for each (input) quantity according to the corresponding probability density function and the total reaction rate calculated from these values is recorded. The procedure is repeated many times until enough samples of the reaction rate have been generated to estimate the properties of the (output) reaction rate probability density function with the required statistical precision. 

Correlations between quantities have to be considered carefully with this technique.
For example, if the strength of a narrow resonance is estimated from a reduced width or a spectroscopic factor, then the uncertainty in the resonance energy enters in the Boltzmann factor of Eq. (\ref{narresrateexpr}) and in the penetration factor of Eq. (\ref{partpartwidth}). Thus the same random value of the resonance energy, drawn from a Gaussian probability density function, must be used in both expressions. 

It was argued in Ref. \cite{Tho99} that if just $5$ resonances contribute to a given reaction rate, where each resonance has $3$ assumed error sources ($E_r$, $\Gamma_p$, $\Gamma_{\gamma}$), and if each of the $15$ independent error sources were sampled  only $10$ times in a Monte Carlo simulation, then the total reaction rate would have to be calculated $10^{15}$ times. Such a procedure, which is not feasible with present-day computers, would provide information about the uncertainty contribution of each input quantity to the total rate. It must be emphasized that we do not attempt to analyze this kind of detailed information here. Instead, our main goal is to find the probability density function for the total reaction rate, which can certainly be achieved with a significantly reduced number of samples (see below).

A computer code, \texttt{RatesMC}, has been written in order to calculate total reaction rates from resonant and nonresonant input parameters using the Monte Carlo technique. For resonances the code computes reaction rates either from the analytical expressions given in Sec. \ref{formalism} or, if required, by numerical integration of Eq. (\ref{generalrate}). The latter procedure, although computationally slow since one integration has to be performed for each randomly drawn set of input quantities, gives the most accurate results if the partial widths of a resonance are known. Upper limits of input quantities and interferences between resonances are also taken into account in the random sampling. The user controls the total number of random samples and hence the precision of the Monte Carlo method. The reaction rate output of the code is described in the next section. Detailed discussions of realistic examples will be given in Paper II, while a description of the input file to \texttt{RatesMC} can be found in Paper III. In Paper IV we compare our new reaction rates to previous results.
%
\subsection{Output statistical distributions of reaction rates}\label{}
\begin{figure}[]
\includegraphics[height=12cm]{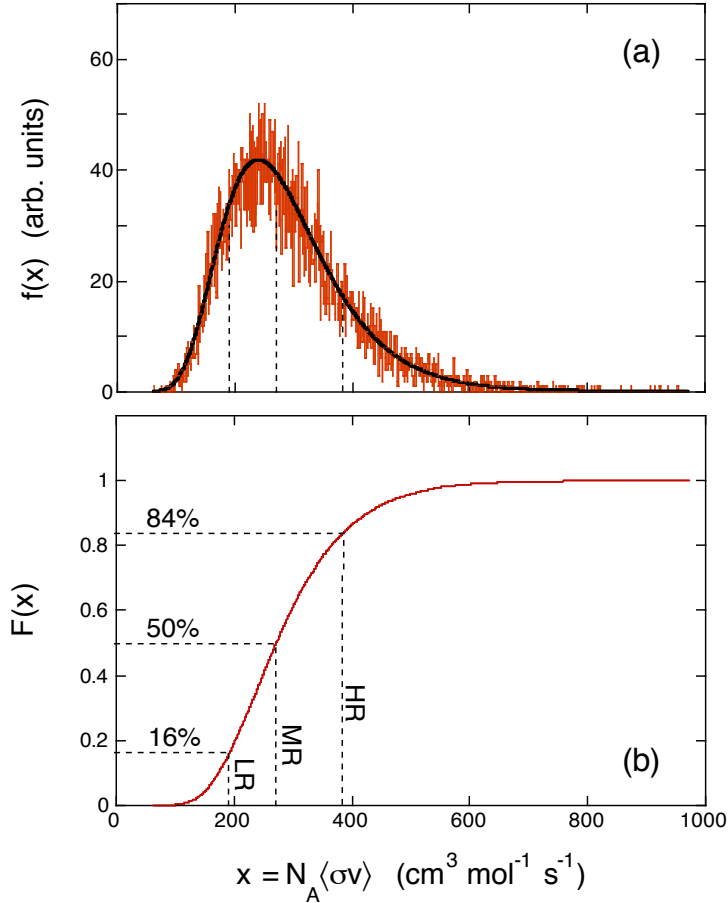}
\caption{\label{fig:sampleMC} 
(Color online) Results of Monte Carlo calculation for a single, fictitious resonance in the $^{22}$Ne($\alpha$,$\gamma$)$^{26}$Mg reaction at a temperature of T=0.5 GK. The resonance parameters are $E_r=300\pm15$~keV and $\omega\gamma=4.1\pm0.2$~eV. The reaction rate is sampled 10000 times. (a) Reaction rate probability density function, shown in red; the black solid line represents a lognormal approximation (see text). (b) Cumulative reaction rate distribution. The vertical dotted lines represent the low, median and high Monte Carlo reaction rates which are obtained from the 0.16, 0.50 and 0.84 quantiles, respectively.}
\end{figure}

Numerical results from the Monte Carlo method, obtained using the code \texttt{RatesMC}, are displayed in Fig. \ref{fig:sampleMC}. They have been obtained for a single, fictitious resonance in the $^{22}$Ne($\alpha$,$\gamma$)$^{26}$Mg reaction at a stellar temperature of T=0.5 GK. The assumed parameters of the narrow resonance are $E_r=300\pm15$~keV and $\omega\gamma=4.1\pm0.2$~eV. The reaction rate has been sampled 10000 times. Part (a) displays the probability density function as a red histogram. The distribution is skewed because of the relatively large uncertainty in the resonance energy (see below). The estimate of the Monte Carlo reaction rate proceeds in the following manner. First, the cumulative rate distribution, shown as a red line in part (b), is calculated from the set of sampled reaction rate values and is normalized to unity. Note the amount of scatter in part (a), while a rather smooth curve is obtained in part (b) with the chosen sample size. Second, the location parameter (or central value), together with a measure for the spread of reaction rate values, are calculated from the cumulative distribution. For the central (recommended) value of the rate we chose the {\it median} which is equal to the 0.50 quantile of the cumulative distribution (Sec. \ref{statdistr}). In our example, this value amounts to $N_A\langle \sigma v \rangle_{med} = 2.72\times10^2$~cm$^3$mol$^{-1}$s$^{-1}$. The low and high values of the rate are chosen to coincide with the 0.16 and 0.84 quantiles. With this choice the confidence level (or the coverage probability) is 68\%. The derived values are $N_A \langle \sigma v \rangle_{low} = 1.92\times10^2$~cm$^3$mol$^{-1}$s$^{-1}$ and $N_A\langle \sigma v \rangle_{high} = 3.87\times10^2$~cm$^3$mol$^{-1}$s$^{-1}$. We will avoid in the following and in Papers II and III expressions such as ``lower limit" or ``maximum rate" since they imply sharp boundaries (see discussion in Sec. \ref{conmean}). Instead we will use the terms {\it low rate} and {\it high rate} when referring to the 0.16 and 0.84 quantiles derived from the Monte Carlo method.

It must be emphasized that we are deriving our results directly from the cumulative distribution\footnote
{In practice, the $N$ sampled values of reaction rates, $x_i$, are sorted into ascending order. The cumulative distribution, $F(x)$, is constant between consecutive values of $x_i$ and rises an equal amount, $1/N$, at each sampled rate. The quantiles (Sec. \ref{statdistr}) are found from the fraction of values located below a given $x_q$, after proper interpolation.} of Monte Carlo reaction rates (Fig. \ref{fig:sampleMC}b) instead of the probability density function (Fig. \ref{fig:sampleMC}a). The reason is that the latter distribution depends on the binning of reaction rate values which introduces considerable arbitrariness as to how the bins should be chosen. Since binning always involves a loss of information, we prefer to derive our numerical results from the cumulative distribution. Plots of reaction rate probability density functions are presented here and in Paper II mainly to aid in the visualization of our results. 

A few comments are in order. An obvious test of our method is the comparison of the value of the median Monte Carlo rate with the classical rate. From Eq. (\ref{narresrateexpr}) one finds $N_A\langle \sigma v \rangle_{class} = 2.71\times10^2$~cm$^3$mol$^{-1}$s$^{-1}$, in agreement with the Monte Carlo result. Such a comparison is of course not possible when upper limits of input quantities need to be considered since the classical reaction rate calculation method does not properly account for upper limits. Therefore, we will consistently quote in Paper II the median rate which represents our recommended reaction rate. In all cases that were analyzed in more detail and that did not involve any upper limits for input quantities we found that the classical rate and the median rate agree within about 5\%. In any case, it is important for a given situation to perform at least two computational runs with a different number of samples in order to test the numerical stability of the results. In most cases, we found that runs with at least 5000 samples were necessary to obtain reaction rates that are reproducible within a few percent.

Recall that the median rate divides the probability density function shown in Fig. \ref{fig:sampleMC}a into two parts of equal area. In this case the {\it median} does not coincide with the {\it mode} (that is, the maximum of the distribution) since the probability density function is skewed. Another measure for the location parameter is the {\it mean} which is obtained from Eq. (\ref{meanvar}). Although in this example the mean agrees numerically with the median, we found in more complicated cases (see Paper II) that the values are not always in agreement. It is a well-known fact that the value of the mean is much more sensitive to outliers in the distribution, while the median in this regard is a more robust measure for the location \cite{Ham93}. Therefore, we prefer to quote the latter over the former quantity.

For reasons that will become clear in the next section it is crucial to find a simple analytical approximation for the Monte Carlo reaction rate probability density function at each temperature. During the evaluation process we have obtained all kinds of shapes for the reaction rate probability density, ranging from highly skewed to symmetric (bell) shapes. A Gaussian is certainly not a reliable approximation to a skewed distribution but, as we have seen in Sec. \ref{lognormal}, a lognormal distribution can account for both symmetric and asymmetric shapes. There are a number of reasons why a lognormal distribution is useful for this purpose. First, suppose that the total reaction rate is dominated by a nonresonant process; since we assume a lognormal probability density function for the effective S-factor, it follows immediately from Eq. (\ref{nonresrateexpr}) that the total reaction rate is also lognormally distributed. Second, if the total reaction rate is dominated by a single resonance and if the uncertainty in the resonance energy dominates over the uncertainty in the resonance strength, which is the case for the situation displayed in Fig. \ref{fig:sampleMC}, then the total rate given by Eq. (\ref{narresrateexpr}) will again be lognormally distributed; this can be seen from the fact that for a Gaussian distributed random variable $y$ (here the resonance energy) the variable $x=e^y$ (here the reaction rate) will follow a lognormal distribution (see Sec. \ref{lognormal}). Third, if the total reaction rate is dominated by a single resonance and if the uncertainty in the resonance strength dominates over the uncertainty in the resonance energy, then the rate distribution will become lognormal since the resonance strength enters linearly in the resonant rate expression; in fact, for a moderate uncertainty in the resonance strength, $\sqrt{V[x]}/E[x]<20$\% (see Sec. \ref{lognormal} and Fig. \ref{fig:GauLog}a), the rate distribution becomes Gaussian in shape. Finally, if the total rate is given by the sum of several contributing resonances, then the total rate will tend toward a Gaussian, according to the central limit theorem. Recall that a Gaussian distribution can be approximated by a lognormal distribution in a straightforward manner. 

These arguments do not prove that the total reaction rate must necessarily be lognormally distributed. In fact, we will demonstrate in Paper II that the Monte Carlo reaction rate probability density functions are in general {\it not} lognormal. However, for the majority of reaction rates that we analyzed in detail, the assumption of a lognormal approximation for the reaction rate probability density function appears useful. Better approximations may exist, for example, the function given by Eq. (\ref{quasilog}) which differs from a lognormal distribution. We have not pursued this idea and further studies may be desirable.

Once a lognormal distribution is adopted for approximating the reaction rate probability density function, it is a simple matter to calculate the lognormal parameters from the expectation value and the variance for $\ln(x)$ (since $\mu = E[\ln(x)]$ and $\sigma^2 = V[\ln(x)]$; see Sec. \ref{lognormal}). For the above example, one finds values of $\mu=5.603$ and $\sigma=0.3526$. The lognormal distribution calculated with these parameters is shown in Fig. \ref{fig:sampleMC}a as a black solid line. It can be seen that the agreement with the Monte Carlo distribution, shown in red, is excellent. Note that the black line does not represent a fit to the data but its parameters are directly derived from the distribution of randomly sampled reaction rates. In Paper II we provide for each temperature point, in addition to the low, median and high Monte Carlo rate, the lognormal parameters $\mu$ and $\sigma$ of the total reaction rate probability density function.

Finally, we comment on a few observations that will prove useful when considering the evaluated reaction rates presented in Paper II. Recall from Sec. \ref{lognormal} that the skewness of a lognormal distribution is related to the magnitude of $\sigma$. As a rule of thumb, values of $\sigma<0.1$ correspond to a nearly symmetric (that is, Gaussian) distribution, while for larger values the distribution will be noticeably skewed. In the above example we found $\sigma=0.3526$, indicating a strongly skewed distribution as is apparent from Fig. \ref{fig:sampleMC}a. Thus a quick inspection of the table columns listing $\sigma$ in Paper II will immediately reveal the skewness of the total reaction rate probability density function. Second, we find that in the majority of cases (that is, when the total reaction rate is lognormally distributed) the lognormal parameters are related to the low, median and high Monte Carlo rates by\footnote{Recall that $x$ is defined to be in units of cm$^3$~mol$^{-1}$~s$^{-1}$. Here, the transformation of $x$ to a dimensionless quantity, for use as an argument of a logarithmic or exponential function, is accomplished implicitly by dividing $x$ by 1 cm$^3$~mol$^{-1}$~s$^{-1}$. Similar transformations apply to $\mu$ and $\sigma$.} (Sec. \ref{lognormal} and Eq. (\ref{geomeanex}))
\begin{equation}
\mu = \ln (x_{med}) = \ln \sqrt{x_{low} x_{high}}~,~~~~~
\sigma = \ln \left(\frac{x_{high}}{x_{med}} \right)= \ln \sqrt{\frac{x_{high}}{x_{low}}} \label{georelation}
\end{equation}
or, equivalently,
\begin{equation}
x_{low}=e^{\mu-\sigma} ~,~~~~~x_{med}=e^{\mu} ~,~~~~~x_{high}=e^{\mu+\sigma}  \label{georelationb}
\end{equation}
where $x$ denotes the total reaction rate. These relationships, which apply to a coverage probability of 68\%, usually hold within a few percent and, in fact, can be used to determine in a simple manner if the total reaction rate probability density function is indeed lognormal or not. They are also useful for estimating an approximate probability density function for some reaction rates that have not been analyzed in the present work with the Monte Carlo technique (see Paper II for details). 
We emphasize that our reaction rate uncertainties reflect our best current knowledge of the nuclear physics input. By no means can our method account for the possibility that, for example, a reported resonance strength was derived using the wrong stoichiometry, or that an incorrect $J^{\pi}$ value has been reported for a particular nuclear level. This issue should be kept in mind when drawing conclusions from our Monte Carlo reaction rate uncertainties.
\subsection{Use of Monte Carlo reaction rates in stellar models}
Finally, the general usefulness of our results for future stellar model simulations will be addressed. It has already been mentioned in Sec. \ref{conmean} that more emphasis has been placed in recent years on studying the influence of reaction rate uncertainties on stellar nucleosynthesis. Consider, for example, the sensitivity study of classical nova nucleosynthesis presented in Ref. \cite{Ili02}. The general strategy adopted by the authors was the following: (i) calculate the temperature-density evolution for specific burning zones within a hydrodynamical model; (ii) set up a grid of ``reaction rate variation factors" within the boundaries provided by published ``lower" and ``upper limits" on the rate; and (iii) perform post-processing reaction network calculations by varying the rate of one reaction at a time. The procedure represents a brute force approach and was intended to provide a first qualitative impression on the reaction rate sensitivity. However, it does not account for the interplay and for correlations between nuclear processes in the reaction network and, even more importantly, it does certainly not provide realistic estimates of isotopic abundances and associated uncertainties for the reasons discussed at length in the present work. 

An interesting, different approach was presented by Hix and collaborators \cite{Hix03}.
These authors assumed (although they did not justify) that reaction rates are distributed according to a lognormal probability density function. Sets of reactions were then grouped according to their global uncertainties: for all reactions involving radioactive targets in their network an uncertainty of a factor 1.5 was assumed; for all stable target nuclei the assumed uncertainty amounted to a factor 1.2, and so on. Note that these ``factor uncertainties" are related to the geometric standard deviation of the lognormal distribution, according to Eq. (\ref{geomeanex}). Finally, post-processing reaction network calculations were performed many times in a Monte Carlo study by simultaneously varying all rates, where for each reaction the rate was randomly sampled according to the corresponding lognormal probability density function. As a result of this procedure, abundance distributions for each nuclide in the network were obtained which could then be analyzed to derive average abundances and associated uncertainties. An obvious drawback of this technique is the fact that {\it global} and rather small reaction rate uncertainties were adopted that had no relationship to any measured or estimated nuclear physics input. 

The present work provides the important information on the reaction rates that was missing so far. In Paper II we report for each reaction in the mass range of interest here the numerical values of the lognormal parameters $\mu$ and $\sigma$ on a temperature grid. These tables can be incorporated into post-processing reaction network calculations. According to Eq. (\ref{georelationb}), the value of $\mu$ provides the median Monte Carlo reaction rate, while the value of $\sigma$ determines the width of the reaction rate probability density function. This is all the information needed in order to perform, in a second Monte Carlo step, a simultaneous variation of all rates in the reaction network. Such investigations will provide more realistic estimates of abundances and their associated uncertainties. We are looking forward to the results of such studies for the nucleosynthesis in red giants, AGB stars, classical novae, supernovae and other sites.
%
\section{Summary and suggestions for future work}\label{summary}
The present work describes a method, based on Monte Carlo techniques, of evaluating  thermonuclear reaction rates. The point is made that reaction rates reported up to now in the literature have no rigorous statistical meaning. As a first step toward a new method, we associate each nuclear physics quantity entering in the calculation of reaction rates with a specific probability density function, including Gaussian, lognormal and chi-squared distributions. Based on these (input) probability density functions the total reaction rate is randomly sampled many times until the required statistical precision is achieved. This procedure results in a median Monte Carlo rate that agrees under certain conditions with the commonly reported recommended ``classical" rate. For each temperature a low rate and a high Monte Carlo rate is computed, corresponding to the 0.16 and 0.84 quantiles of the cumulative reaction rate distribution. These quantities differ in general from the statistically meaningless ``minimum" (or ``lower limit") and ``maximum" (or ``upper limit") reaction rates which are commonly reported in the literature. In addition, we approximate the (output) reaction rate probability density function by a lognormal distribution and present, at each temperature, the lognormal parameters $\mu$ and $\sigma$. The values of these quantities will be important in future Monte Carlo nucleosynthesis studies. Our new reaction rates, appropriate for {\it bare nuclei in the laboratory}, are tabulated in the second paper of this series (Paper II). The nuclear physics input used to compute the reaction rates, together with a description of the Monte Carlo code \texttt{RatesMC}, is presented in the third paper (Paper III). In the fourth paper (Paper IV) we compare our new reaction rates to previous results.

We summarize below certain aspects of our work that call for future efforts from the nuclear astrophysics community: \\
(1) We can hardly overemphasize that incomplete information is usually published when  the results of a measurement are reported. It is not sufficient to provide a value and its standard  deviation, but the probability density function on which these values are based should also be reported. This is especially important for null-results: to report simply an ``upper limit" together with a confidence level is insufficient, unless the most important piece of information, that is, the corresponding probability density function, is reported as well.\\
(2) Null-results are incorporated in a consistent way into the present Monte Carlo method within the framework of Porter-Thomas distributions. In order to draw a random sample of a reduced width (or a spectroscopic factor) from a Porter-Thomas distribution, the {\it local mean value} must be known. It is reasonable to assume that this mean depends on the nuclear mass number $A$ and the orbital angular momentum $\ell$. In the present work we use values for the mean that have been obtained from our preliminary analysis, that is, by binning together all values of a large data sample and fitting them to a Porter-Thomas distribution, regardless of their $A$ or $\ell$ values. What is required are systematic studies of nuclear statistical properties that provide improved local mean values for proton and $\alpha$-particle reduced widths. Similar studies should be performed for reduced $\gamma$-ray transition probabilities. Theoretical investigations, perhaps employing the shell-model, could be helpful in this regard.\\ 
(3) We approximate the output reaction rate probability density function by a lognormal distribution and provide reasonable arguments for justifying this assumption. However, in some cases, especially when the uncertainty on the resonance energy is large or when undetected low-energy resonances become important, the output reaction rate distributions deviate strongly from lognormality. In such cases we obtain results which can only be approximated by a statistical distribution that depends on more than two parameters. Further studies are required to decide if more complicated expressions of reaction rate probability density functions are needed for future nucleosynthesis studies.\\
%
\section{Acknowledgement}
The authors would like to thank Gary Mitchell for helpful discussions. This work was supported in part by the U.S. Department of Energy under Contract No. DE-FG02-97ER41041. 



\end{document}